\documentstyle[12pt, a4]{article}
\begin{document}
\hyphenation{anti-fermion}
\topmargin= -20mm
\textheight= 230mm
\textwidth= 8.5 in
\baselineskip = 0.33 in

 \begin{center}
\begin{large}
 {\bf {  FINITE SIZE CORRECTIONS 
 
 IN MASSIVE THIRRING MODEL }}

\end{large}

\vspace{1cm}

T. FUJITA\footnote{e-mail: fffujita@phys.cst.nihon-u.ac.jp}
 and H. TAKAHASHI\footnote{e-mail: htaka@phys.cst.nihon-u.ac.jp}

Department of Physics, Faculty of Science and Technology  
  
Nihon University, Tokyo, Japan

\vspace{1cm}

{\large ABSTRACT} 

\end{center}

We calculate for the first time 
 the finite size corrections in the massive Thirring model. 
This is done by numerically solving the equations of  periodic boundary 
conditions of the Bethe ansatz solution.  
It is found  that the corresponding central charge 
extracted from the $1/L$ term is around 0.4 for 
the  coupling constant of ${g_0}=-{\pi\over 4} $ and decreases 
down to zero when ${g_0}=-{\pi\over{3}}$. 
This is 
quite different from the predicted central charge of the sine-Gordon model.

\vspace{1cm}
  \par

\vspace{2cm}

\newpage



In two dimensional field theory, there is a remarkable correspondence 
between the fermionic and bosonic field theories. This was 
first recognized by Coleman [1], and he proved that the sine-Gordon 
field theory and the massive Thirring model are equivalent 
to each other in that 
the arbitrary order of the correlation functions turn out to be the same. 

Recently, however, Klassen and Melzer [2] argue that the equivalence 
between the sine-Gordon and the massive Thirring models may be 
violated at the finite size correction. They proved 
by using the perturbed conformal field theory that these two models 
are different in finite-volume  energy levels, for example. 

In this paper, we  calculate the finite size corrections to the 
ground state energy. We solve numerically the equations of 
the periodic boundary condition  
in the Bethe ansatz solutions of the massive Thirring model [3-5]. The ground 
state energy can be expressed as 
$$ E_{v} = E_0 L - {\pi \tilde{c} \over{6L}} + ...  \eqno{(1)} $$
where $L$ denotes the box size.  $\tilde{c}$ 
corresponds to a central charge at the massless limit [6,7]. 

The present calculation shows that the corresponding 
central charge $\tilde{c}$ 
in the negative coupling constant regions (no bound states) is around $0.4$ 
for  ${g_0}=-{\pi\over 4}$ and that it 
becomes zero when ${g_0} = -{\pi\over {3}}$. These values can be 
compared with those calculated for the sine-Gordon field theory [8,9].  
The central charge for the sine-Gordon 
field theory with the massless limit can be expressed as 
$$ c=1-{6\over{p(p+1)}} \eqno{(2)}$$ 
where $p$ is an integer  
and is related to the coupling constant $g_0$ as 
$$ {g_0}=-{\pi\over 2}(1-{1\over{p}})  . \eqno{(3)} $$ 
In fig.1, we summarize the calculated central charge as the function of 
the coupling constant for the sine-Gordon model by Itoyama and Moxhay, 
and for the massive Thirring model by the present calculations. One can see 
that the values of the central charge predicted for the two models
 are very different from each other. 

It is, however, not very clear to us  
whether this difference may be related to 
a possible violation of  the equivalence 
between the sine-Gordon field theory and the massive Thirring model 
at the finite volume energy as suggested by Klassen and Melzer. 

\vspace{1cm}


Here, we briefly review  
the massive Thirring model whose lagrangian density can be written as [10]
$$  {\cal L} =  \bar \psi ( i \gamma_{\mu} \partial^{\mu} - m_0 ) \psi 
  -{1\over 2} g_0 j^{\mu} j_{\mu}   \eqno{(4)} $$
with the fermion current $j_{\mu} = :\bar \psi  \gamma_{\mu} \psi : $. 
Choosing a basis where $\gamma_5$ is diagonal, 
we write the hamiltonian as 
$$  H = \int dx \left[-i(\psi_1^{\dagger}{\partial\over{\partial x}}\psi_1
-\psi_2^{\dagger}{\partial\over{\partial x}}\psi_2 )+
m_0(\psi_1^{\dagger}\psi_2+\psi_2^{\dagger}\psi_1 )+
2g_0 \psi_1^{\dagger}\psi_2^{\dagger}\psi_2\psi_1 \right]  .  \eqno{(5)} $$

The hamiltonian eq.(5) can be diagonalized by the Bethe ansatz wave 
functions $ \Psi (x_1,...,x_N)$ 
with $N$ particles 
$$ \Psi (x_1,...,x_N)= \exp(im_0 \sum x_i \sinh \beta_i ) 
\prod_{1\leq i < j \leq N} \left[1+i\lambda (\beta_i,\beta_j) \epsilon (x_i-x_j) \right] 
\eqno{(6)} $$
where $\beta_i$ is related to the momentum $k_i$ 
 and the energy $E_i$ of $i$-th particle as 
$$ k_i= m_0 \sinh \beta_i   .  \eqno{(7a)} $$
$$ E_i= m_0 \cosh \beta_i   .  \eqno{(7b)} $$
where $\beta_i$'s are complex variables. 

$ \epsilon (x)$ is a step function and is defined as 
$$ 
\epsilon (x) =  
\left \{
\begin{array}{rc}
-1 & x<0 \\
\ \ \\
1 & x>0 .
\end{array}
\right .
    \eqno{(8)} $$ 
$ \lambda (\beta_i,\beta_j)$ is related to the phase shift function 
$ \phi (\beta_i-\beta_j)$ as 
$$  { {1+i \lambda (\beta_i,\beta_j)}\over{1-i \lambda (\beta_i,\beta_j)}}=
\exp \left( i \phi (\beta_i-\beta_j) \right) .\eqno{(9)} $$ 
The phase shift function $\phi (\beta_i-\beta_j)$ can be explicitly written as 
$$ \phi (\beta_i-\beta_j)= -2\tan^{-1} \left[{1\over 2}g_0 \tanh {1\over 2}
(\beta_i-\beta_j) \right]    . \eqno{(10)} $$ 

From the definition of the rapidity variable $\beta_i$'s, one sees that 
for positive energy particles, $\beta_i$'s are real while for 
negative energy particles, $\beta_i$ takes the form $i\pi -\alpha_i$ 
where $\alpha_i$'s are real.  


Since the Bethe ansatz wave functions diagonalize the hamiltonian, 
we demand that they satisfy the periodic boundary conditions (PBC) 
with the box length $L$ [3],  
$$ \Psi (x_i=0) = \Psi (x_i=L) .  \eqno{(11)} $$
This leads to the following PBC equations, 
$$ m_0 L \sinh \beta_i = {2\pi n_i} - \sum_j \phi (\beta_i -\beta_j)  
      \eqno{(12)} $$ 
where $n_i$'s are integer. Here, we note that we cannot take the 
anti-periodic boundary condition since it does not reproduce the boson 
spectrum in the positive coupling constant regions [5].   

\vspace{1cm}

The parameters we have here 
are the box length $L$ and the particle number $N$. 
 In this case, the density of the system $\rho$ becomes  
$$ \rho = {N \over{L}}    .  \eqno{(13)}  $$ 
Here, the system is fully characterized by
the density $\rho$. 


We write the PBC equations 
for the vacuum which is filled with negative energy particles 
( $\beta_i=i\pi -\alpha_i$ ),  
$$  \sinh \alpha_i = {2\pi n_i \over{L_0}}  
 - {2\over{L_0}} \sum_{j \not= i} \tan^{-1}\left[{1\over 2}g_0 
\tanh{1\over 2} (\alpha_i -\alpha_j) \right] , 
    \eqno{(14)} $$ 
where $ n_i = 0, \pm 1, \pm 2, ..., \pm N_0    $ with 
$N_0 ={1\over 2}(N-1)$ and $L_0=m_0 L$.   

In this case, the vacuum energy $E_v$ can be written as  
$$ E_v =- \sum_{i=-N_0}^{N_0} m_0 \cosh \alpha_i  . \eqno{(15)} $$ 

In this paper, we have carried out the numerical calculations of the PBC 
equations. The numerical method to solve the PBC equations is 
explained in detail in ref.[5]. 

Now, the calculated vacuum energy can be parametrized as 
$$ E_v = E_0 L- {\pi \tilde{c}(g_0)\over{6L}} + ...  \eqno{(16)}  $$
where $\tilde{c}(g_0)$ corresponds 
to the central charge at the massless limit. In what follows, we call 
this $\tilde{c}(g_0)$ as the central charge even though we are solving  
the massive field theory. 
It should be noted that the first term in eq.(16) can be evaluated 
analytically by taking the thermodynamic limit [3]. 

Since we can vary the values of $L$ and $N$, we obtain  
 the corresponding central charge $\tilde{c}(g_0)$. 
 Although we have still rather small 
particle number ($N \sim 10000$), we believe that the values extracted 
for the central charge must be reasonably reliable. 

Now, we want to obtain  the central charge $\tilde{c}(g_0)$ 
at the field theory limit $\rho \rightarrow \infty$. 
In fig. 2, we show the calculated central charge $\tilde{c}(g_0)$ 
as the function 
of the effective density $\rho_0 = {N_0\over{L_0}}$. It is quite 
interesting to observe that the calculated central charge can be well 
parametrized by the following simple formula [11],
$$ \tilde{c}(g_0)= A  + 
B \exp \left( -{\kappa\over{\rho_0}} \right)  \eqno{(17)} $$
where $A$, $B$ and $\kappa$ are constants. 
Therefore, the field theory limit 
can be easily taken since we can let $\rho_0$ infinity.

In Table 1, we show the values of $A$, $B$ and $\kappa $ 
for some values of the coupling constant ${g_0} $.  
The central charge becomes $A+B$ at the field theory limit. 
The calculated values of the central charge are shown 
as the function of the coupling constant ${g_0}$ in fig.1. 
We also plot the central charge calculated for the sine-Gordon 
theory by Itoyama and Moxhay [9]. 
As can be seen from the fig.1, the two values of the central 
charge are quite different from each other. 

\vspace{1cm}

How can we interprete these differences ? 
The first possibility is that the two theories (sine-Gordon and 
massive Thirring models) are different from each other at the 
finite volume. We do not know whether this difference 
can show up as the central charge or not. However, 
a simple-minded physical intuition suggests that the central charge 
which should correspond to the heat capacity cannot be different 
if all the correlation functions of the two models 
are the same with each other. In this case, we should rather check 
the convergence of the perturbation expansions in  
 Coleman's proof of the equivalence between the sine-Gordon 
and the massive Thirring models since it crucially depends on the convergence 
of the expansions. For the negative values of the coupling constant, 
we do not know whether this convergence is already verified or not. 

The second possibility is that neither of the calculations 
are accurate enough to argue the difference between them. 
To this, we should comment on the accuracy of the present 
calculations. Since we have only the limited number of particles, 
we always face the criticism that the real nature (even though 1+1 
dimension) must be 
with the infinite number of particles. We have varied 
the number of particles from 1000 to 10000. It seems to us that 
the extracted central charge may well be reliable to within 
a few tens of percents. At least, we believe that the calculation 
must be rather reliable for the coupling constant around ${g_0} 
= -{\pi\over 4}$ where the extracted central charge is not very small. 
On the other hand, the present calculation may involve 
somewhat large errors 
for the coupling constant around or smaller than ${g_0} 
= -{\pi\over {3}}$ since the extracted central charge is rather small. 
This is in contrast to the bound state problems [5,12-13] 
where there is some possibility of controlling the accuracy 
of the numerical calculations. 
However, the evaluation of the central charge involves rather complicated 
processes of extracting it since we have to obtain it from the term 
proportional to $1\over L$ in the vacuum energy. 
Therefore, the error bars of the calculations we have shown in fig.1 
may well be still optimistic numbers.  

Concerning the central charge of the sine-Gordon model, we do not know 
whether the central charge predicted by Itoyama and Moxhay can be taken to 
be exact or not. Here, we only make a comment on the $string$ hypothesis 
in the massive Thirring model when they employ the thermodynamic 
Bethe ansatz [14]. As discussed in ref.[5,11], the $string$ 
picture in the massive Thirring model in the positive values of the 
coupling constant turns out to be invalid in the sense that they do not 
satisfy the PBC equations. However, in the negative values of the 
coupling constant, we do not know whether there is a $string-$like 
solution that satisfies the PBC equations. 

\vspace{1cm}

We would like to thank M. Hiramoto, C. Itoi,  M. Kato and 
H. Mukaida for helpful discussions and comments.

\vspace{3cm}

\begin{table}[h]

\vspace{1mm}
\begin{center}
Table 1
\vspace{0.5cm}

\begin{tabular}{c|ccc|c} \hline 
\rule[-4mm]{0pt}{4mm} \rule{0pt}{5mm}
   $\frac{g_0}{\pi}$ & $A$  &  $B$ & $\kappa$ & $\tilde{c}(g_0)$  \\ \hline
 \rule{0pt}{5mm}  
 - 1/4    &  0.941 & - 0.562 & 25   & 0.38 $\pm$ 0.09 \\
 - 0.276  &  1.06  & - 0.745 & 16.5 & 0.32 $\pm$ 0.05 \\
 - 0.291  &  1.06  & - 0.795 & 16   & 0.27 $\pm$ 0.06 \\
 - 0.305  &  0.901 & - 0.739 & 25   & 0.16 $\pm$ 0.07 \\
 - 0.319  &  0.854 & - 0.790 & 30   & 0.06 $\pm$ 0.04 \\ 
 - 1/3    &  0.793 & - 0.879 & 40   & - 0.09 $\pm$ 0.10  
  \rule[-3mm]{0pt}{5mm} \\ \hline
\end{tabular}
\label{table_cg}
\end{center}
\vspace{0.5cm}
\begin{center}
We show the 
values of $A$, $B$ and $\kappa$ for some coupling constant $g_0$ 

together
with the $\tilde{c}(g_0)$  at the field theory 
limit.
\end{center}

\end{table}

\newpage

{\large REFERENCES }
\baselineskip = 8 mm

1. S. Coleman, Phys. Rev. {\bf D11}, 2088 (1975) 

2. T.R. Klassen and E. Melzer, Int. J. Mod. Phys. {\bf A8}, 4131 (1993)

3. H. Bergknoff and H.B. Thacker, Phys. Rev. Lett. {\bf 42}, 135 (1979)  

4. H.B. Thacker, Rev. Mod. Phys. {\bf 53}, 253 (1981) 

5. T. Fujita, Y. Sekiguchi and K. Yamamoto, Ann. Phys. {\bf 255}, 204 (1997)

6. H.W. Bl\"ote, J.L. Cardy and M.P. Nightingale, Phys. Rev. Lett. 
{\bf 56}, 742 (1986)

7. I. Affleck, Phys. Rev. Lett. {\bf 56}, 746 (1986) 

8. T. Eguchi and S.K. Yang, Phys. Lett. {\bf B 230}, 373 (1989)

9. H. Itoyama and P. Moxhay, Phys. Rev. Lett. {\bf 65}, 2102 (1990)

10. W. Thirring, Ann. Phys. {\bf 3}, 91 (1958)

11. T. Fujita and M. Hiramoto, NUP-A-97-12, hep-th/9705140

12. T. Fujita and A. Ogura,  Prog. Theor. Phys. {\bf 89}, 23 (1993) 

13. A. Ogura, T. Tomachi and T. Fujita, Ann. Phys. (N.Y.) 
{\bf 237}, 12 (1995)

14. A.B. Zamolodchikov, Nucl. Phys. {\bf B 342}, 695 (1990)

\vspace{1cm}

%
%









\end{document}